# Pi Fractions for Generating Uniformly Distributed Sampling Points in Global Search and Optimization Algorithms

*(Updated 17 March 2014)*

Richard A. Formato[1]

***Abstract:*** π Fractions are used to create deterministic uniformly distributed pseudorandom decision space sample points for a global search and optimization algorithm. These fractions appear to be uniformly distributed on [0,1] and can be used in any stochastic algorithm rendering it effectively deterministic without compromising its ability to explore the decision space. π Fractions are generated using the BBP π digit extraction algorithm. The π Fraction approach is tested using genetic algorithm πGASR with very good results. A π Fraction data file is available upon request.

## 1. Introduction

The uniformity of randomly generated sample points is an important consideration in global search and optimization. Sample points should be generated using a truly uniformly distributed random variable calculated from a probability distribution, but most pseudorandom sequence generators fall short because their points in fact are not uniformly distributed. One alternative approach is using Low Discrepancy Sequences (LDS) in which there appears to be growing interest. De Rainville *et al*. [1,2] provide a summary of the uniformity problem and develop an evolutionary optimization approach to generating LDS. Pant *et al*. [3] describe an improved Particle Swarm Optimization (PSO) algorithm utilizing van der Corput and Sobol LDS. Other representative, not exhaustive examples include LDS applied to liquid crystal display dot patterns [4], power system stabilizers [5], and financial analysis [6,7].

This note describes another alternative for generating uniformly distributed sample points using π Fractions computed from hexadecimal digit extraction from the constant π. Pi Fractions are uniformly distributed and provide a basis for creating reproducible sample point distributions that can be used in any global search and optimization algorithm regardless of its nature, stochastic, deterministic or hybrid.

## 2. BBP Algorithm

The Bailey-Borwein-Plouffe (BBP) algorithm quite remarkably extracts hexadecimal digits from the numerical constant π beginning at *any* digit without having to compute *any* of the preceding digits. BBP is based on the identity

$$\pi = \sum_{k=0}^{\infty} \frac{1}{16^k}\left(\frac{4}{8k+1} - \frac{2}{8k+4} - \frac{1}{8k+5} - \frac{1}{8k+6}\right) \quad (1)$$

whose derivation and use in BBP is described in detail in [8]. As an example, the hex digits of π starting at digit 1,000,000 are 26C65E52CB459350050E4BB1 and the corresponding π Fraction is 0.151464362347971272412488292131. For all practical purposes the first 215,829 π Fractions are uniformly distributed on [0,1] with a mean value of 0.499283729688375. The Cumulative Distribution Function (CDF) for these data is plotted in Fig. 1 (1000 bins) in which $\Pr\{\pi_i \leq X\}$ is the probability that $\pi$ Fraction $\pi_i$ is less than or equal to $0 \leq X \leq 1$. It is reasonable to speculate that all sequential $\pi_i$ are uniformly distributed. Testing on various subsets of this data set reveals a uniform distribution regardless of how many contiguous fractions are included in any fairly large sample or where the sequence is begun. It also seems reasonable to believe that any sufficiently large set of arbitrarily selected $\pi_i$ also will be uniformly distributed in [0,1]. These characteristics have not been investigated for the other constants discussed in [8], but they well may exhibit similar behavior.

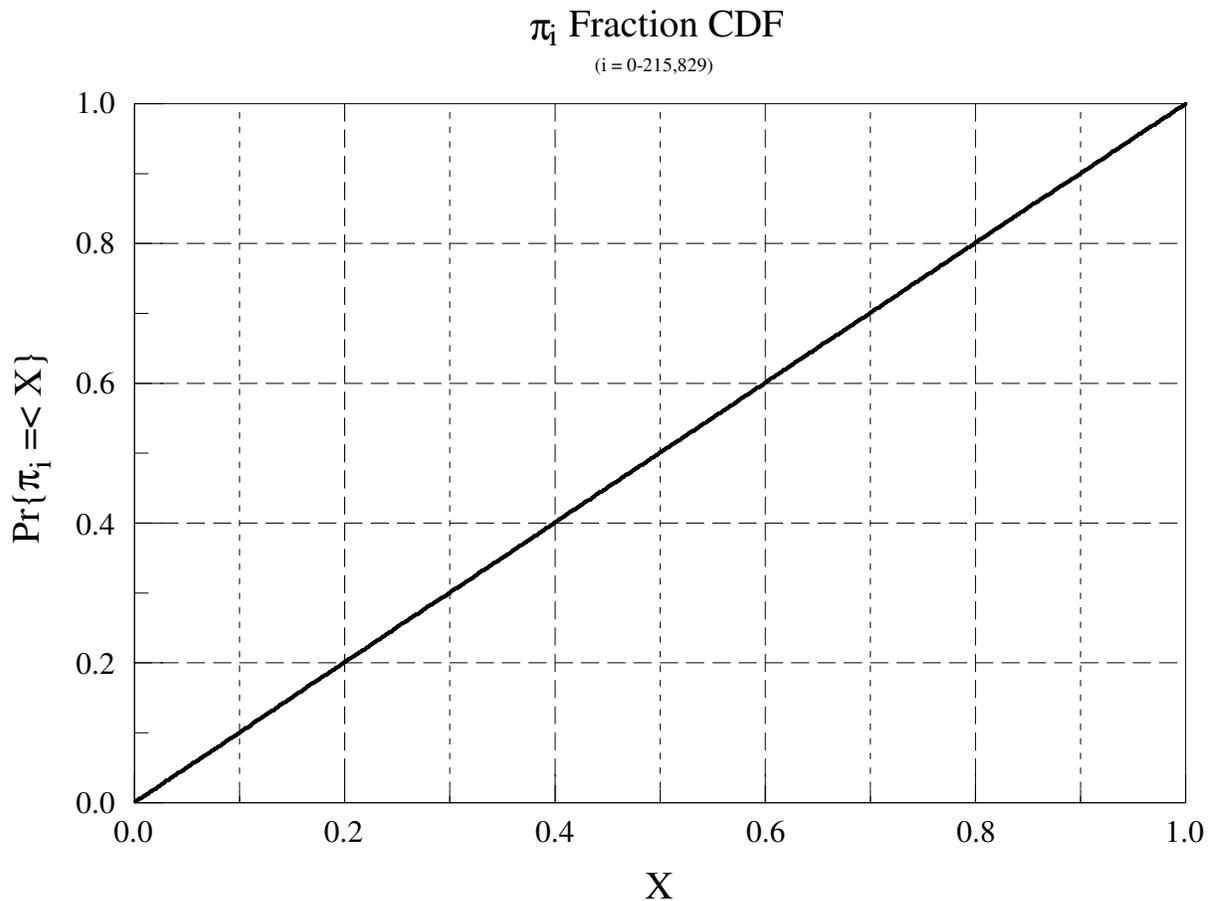

Figure 1. $\pi$ Fraction Cumulative Distribution Function.

### 3. Dimensional Correlations

Nonuniformity in LDS sequences often is evident in bidimenional plots in high dimensionality spaces. Figures 2 and 3 in [1] are good examples. They show, respectively, almost perfect linear correlations in monotonically increasing van der Corput sequences and correlations between dimensions 7 and 8 in a Halton sequence. Other striking visual examples appear in [6] and [7]. Testing of van der Corput and Halton sequences reveals many undesirable correlations. A typical 30-dimensional Halton example for coordinates 27 and 28 appears Fig. 2.

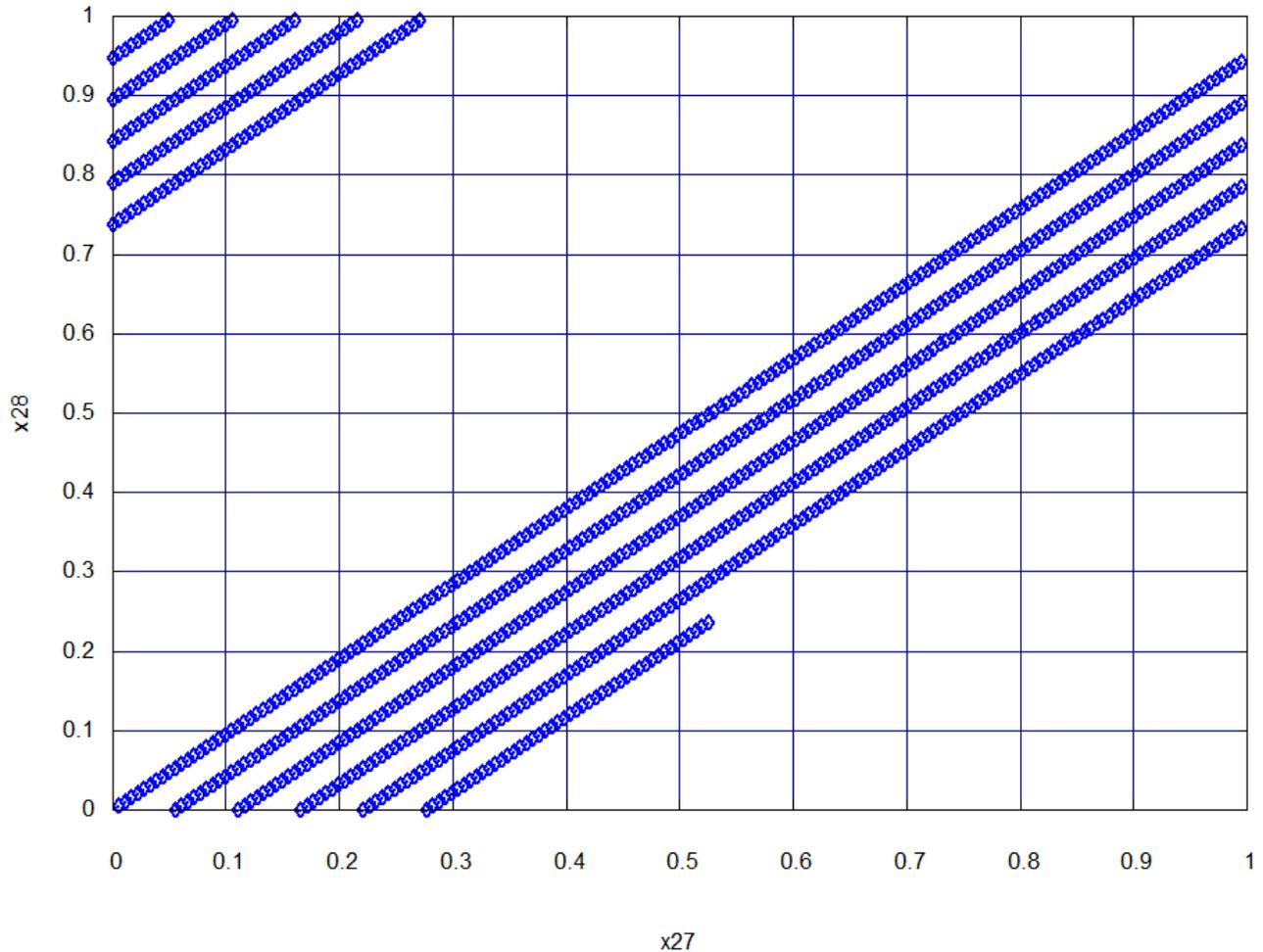

Figure 2. Coordinates ($x_{27}, x_{28}$) 30D Halton sequence (1000 points).

Sample points based on π Fractions also can exhibit strong linear correlations, but apparently only under very limited circumstances. For example, Fig. 3 plots ($x_{27}, x_{28}$) for 1,000 points in 30 dimensions using the π Fractions in their order of occurrence (index increment = 1 starting with the first π Fraction ). The linear correlation is obvious, but it disappears completely when dimensions 27 and 29 are compared instead as seen in Fig. 4. Many test runs suggest that the π Fractions exhibit correlation *only* in successive dimensions and *only* when accessed in their order of occurrence, regardless of where the sequence is started. But when a different index greater than 1 is used, for example, a value of 2, there is no obvious correlation as the ($x_{27}, x_{28}$) plot in Fig. 5 shows. These data suggest it is reasonable to believe that the π Fractions indeed do provide uniformly distributed uncorrelated sample points as long as successive fractions are *not* used to compute the sample point coordinates. A source code listing for the test program that created the correlation plots appears in Appendix 1.

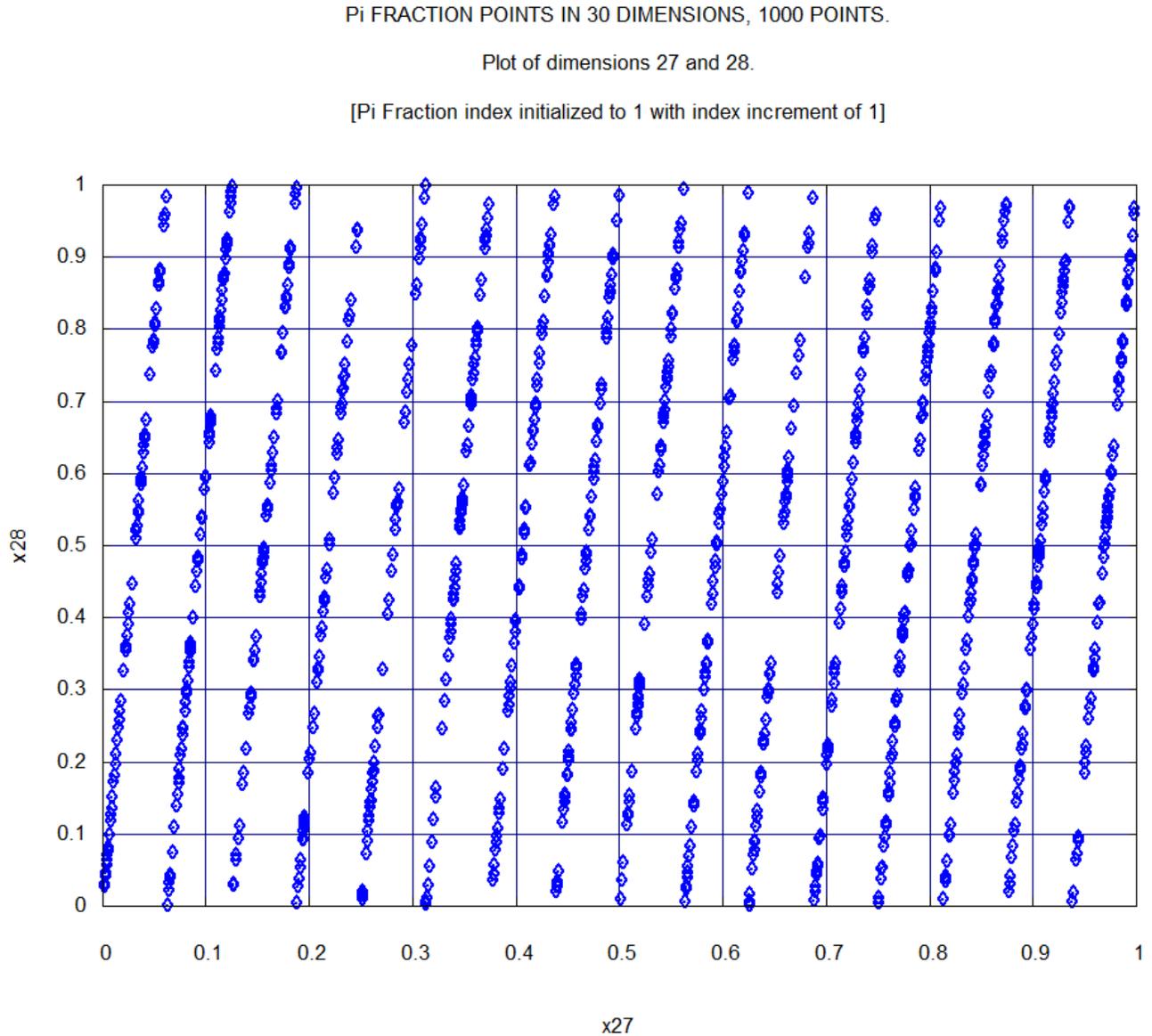

Figure 3. Coordinates ($x_{27}, x_{28}$) 30D $\pi$ Fractions, index increment = 1 (1000 points).

4. πGASR Algorithm

The utility of π Fractions was investigated by generating uniformly distributed sample points and pseudorandom numbers in the genetic algorithm πGASR which is based on Li *et al.*'s novel GA [9]. A standard GA is improved in [9] by (i) allowing competition between child chromosomes in a new crossover operator resulting in better interpolation and extrapolation of decision space sample points and (ii) introducing an iteration-dependent mutation operator. Li *et*

*al.*'s algorithm is referred to here as "Genetic Algorithm with Sibling Rivalry" (GASR) because of the new crossover operator (see [9] for details and note that the "SR" descriptor is introduced here). Its implementation using π Fractions is algorithm πGASR (source code listing in Appendix 2). π Fractions are used to create the initial chromosome distribution and in testing for crossover, mutation, and elitism. Details of the scheme selecting the $\pi_i$ are determined by the algorithm designer and in this case appear in the source code listing. Note that the manner in which the πGASR's fractions are selected avoids the bidimensional correlation issue in §3. Note too that πGASR maximizes the objective function instead of minimizing it.

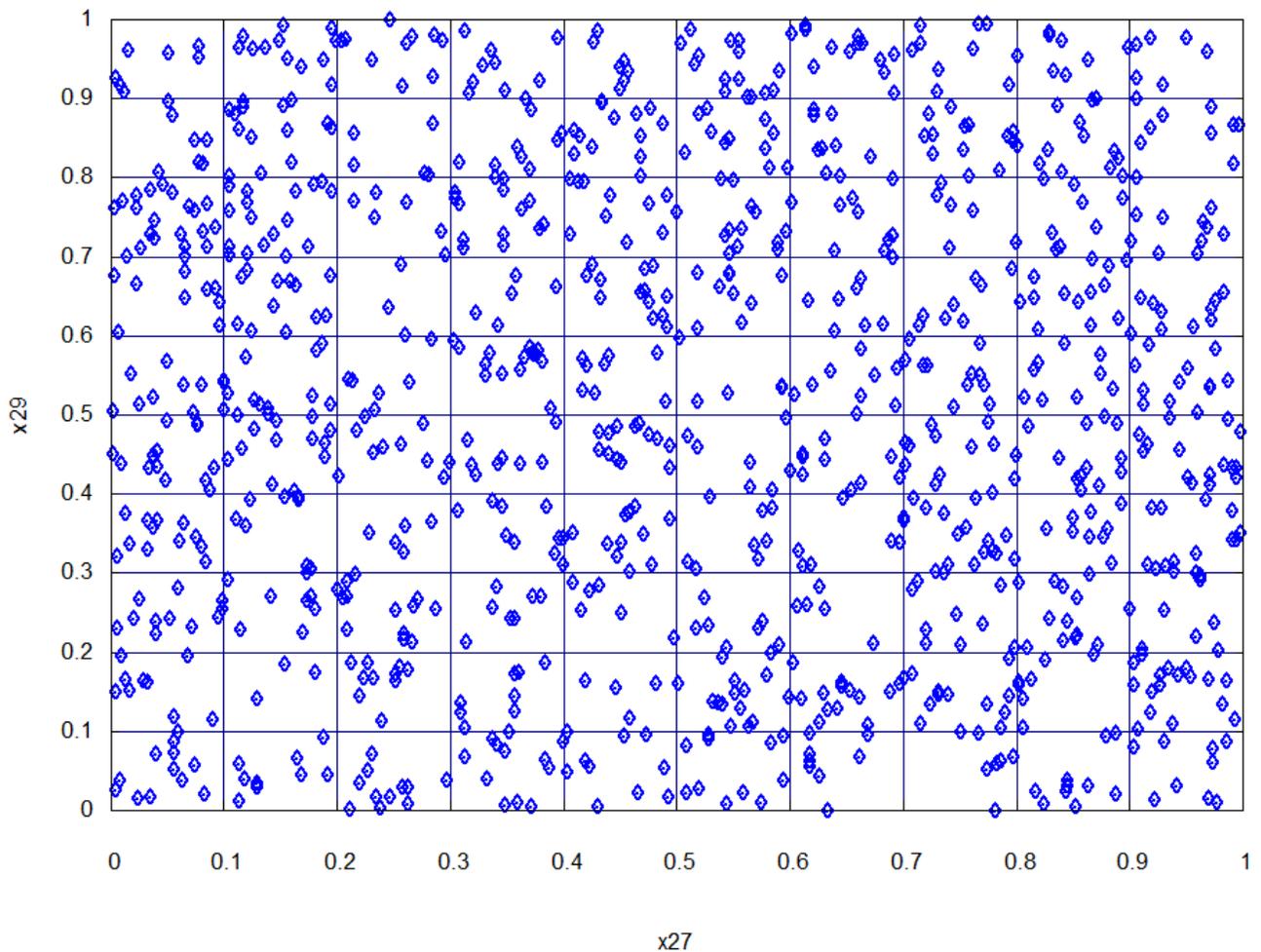

Figure 4. Coordinates $(x_{27}, x_{29})$ 30D π Fractions, index increment = 1 (1000 points).

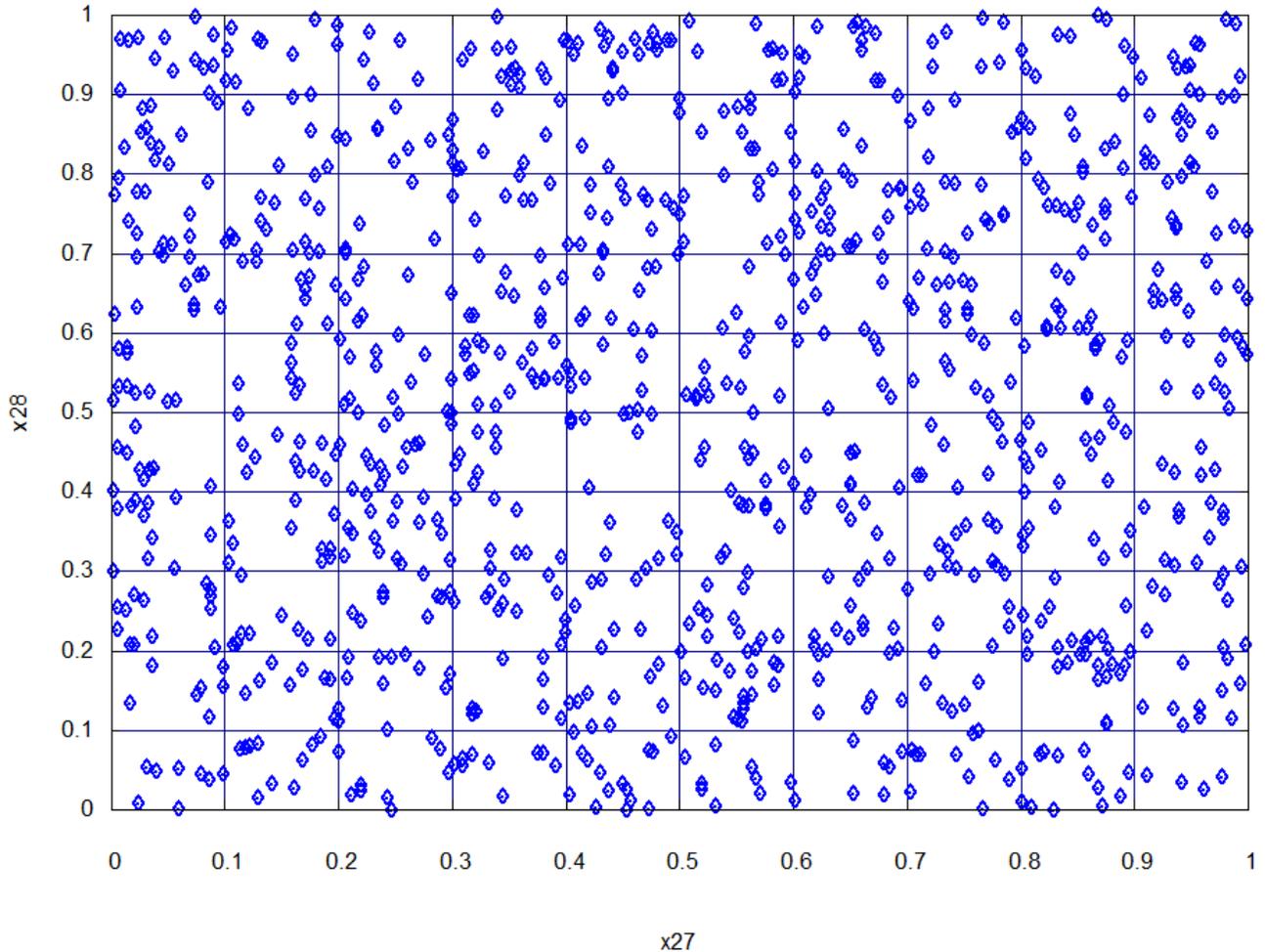

Figure 5. Coordinates ($x_{27}, x_{28}$) 30D π Fractions, index increment = 2 (1000 points).

## 5. Benchmark Results

πGASR was tested against the six-function benchmark suite shown in Table 1. DS is the decision space, $x*$ the location of the objective function's known maximum, and $f(x*)$ its value. This suite was used in [10] to test the new algorithm vibrational-PSO (v-PSO). Table 2 compares v-PSO and πGASR results for 10, 20 and 30-dimensional benchmarks. $N_d$ is the DS dimensionality and $N_{eval}$ the total number of πGASR function evaluations. The v-PSO data are average values for 100 runs using 200,000 function evaluations per run (20,000,000 evaluations of each test function). Because v-PSO performs minimization the signs of its results have been changed for comparison to πGASR. In all cases, a single πGASR run was made. The algorithm's performance likely will be much better if many runs are made with different π

Fraction distributions, but that was not investigated because πGASR performed quite well considering the relatively small number of function evaluations. In addition, every πGASR run with specific π Fraction distributions yields the same result every time because the fractions are pseudorandom and therefore known with absolute precision.

In terms of function evaluations, πGASR's worst case figure of 656,308 is nearly *97% less* than v-PSO's 20,000,000. In terms of solution quality, πGASR performed very well on $f_2$, $f_3$ and $f_4$ ; well on $f_1$ and $f_5$ ; and exceptionally well on $f_6$. For $f_6$ with $N_d$=30 πGASR required 294,763 evaluations (98.5% fewer than v-PSO) and returned a best fitness of $-9.400238 \times 10^{-3}$ compared to v-PSO's average value of -2,139.5±103.3. These results strongly suggest that pseudorandom π Fractions can be very useful in implementing what amount to *deterministic* "stochastic" algorithms.

## 6. Conclusion

π Fractions have been shown to be an effective approach to creating uniformly distributed decision space sample points for global search and optimization. Fractions associated with constants other than π may be similarly useful. Algorithm πGASR was used an example, and its performance tested against the v-PSO six benchmark suite with generally very good results and in one case much better results. π Fraction pseudorandom sequences should be useful for improving the performance of any "stochastic" algorithm in several ways: (i) the resulting sequences are entirely deterministic so that all runs with the same setup produce exactly the same results thus rendering a stochastic algorithm effectively deterministic without compromising its ability to explore the decision space; (ii) making successive runs with different sequences likely will result in better performance with far fewer function evaluations; and (iii) decision space adaptation is easily accomplished because the sequences are deterministic (for example, shrinking the decision space around a group of maxima). A π Fraction data file and source code listings are available from the author; email requests to `rf2@ieee.org`.

[1]Consulting Engineer &
 Registered Patent Attorney
 *Of Counsel*, emeritus
 Cataldo & Fisher, LLC
 P.O. Box 1714
 Harwich, MA 02645 USA
 rf2@ieee.org

*Saint Augustine, Florida*
*March 17, 2014*

## Table 1. v-PSO Benchmark Functions.

| Fnc# | Function | $f(x)$ | DS | $x^*$ | $f(x^*)$ |
|---|---|---|---|---|---|
| $f_1$ | Ackley | $20\exp\left(-0.2\sqrt{\frac{1}{N_d}\sum_{i=1}^{N_d}x_i^2}\right) + \exp\left(\frac{1}{N_d}\sum_{i=1}^{N_d}\cos(2\pi x_i)\right) - 20 - e$ | $[-30,30]^{N_d}$ | $[0]^{N_d}$ | 0 |
| $f_2$ | Cosine Mixture | $-\sum_{i=1}^{N_d}x_i^2 + 0.1\sum_{i=1}^{N_d}\cos(5\pi x_i)$ | $[-1,1]^{N_d}$ | $[0]^{N_d}$ | $0.1 N_d$ |
| $f_3$ | Exponential | $\exp(-0.5\sum_{i=1}^{N_d}x_i^2)$ | $[-1,1]^{N_d}$ | $[0]^{N_d}$ | 1 |
| $f_4$ | Griewank | $-\frac{1}{4000}\sum_{i=1}^{N_d}(x_i-100)^2 + \prod_{i=1}^{N_d}\cos\left(\frac{x_i-100}{\sqrt{i}}\right) - 1$ | $[-600,600]^{N_d}$ | $[0]^{N_d}$ | 0 |
| $f_5$ | Rastrigin | $-\sum_{i=1}^{N_d}[x_i^2 - 10\cos(2\pi x_i) + 10]$ | $[-5.12,5.12]^{N_d}$ | $[0]^{N_d}$ | 0 |
| $f_6$ | Schwefel | $-418.9829 N_d + \sum_{i=1}^{N_d}[x_i \sin(\sqrt{|x_i|})]$ | $[-500,500]^{N_d}$ | $[420.9687]^{N_d}$ | 0 |

## Table 2. πGASR Results for v-PSO Benchmark Suite.

| fnc# | $N_d$ | Best Fitness v-PSO* | Best Fitness πGASR | Neval |
|---|---|---|---|---|
| $f_1$ | 10 | –1.84e-15±2.9e-16 | -5.762878e-4 | 656,308 |
|  | 20 | –2.84e-15±1.5e-16 | -1.161337e-2 | 328,243 |
|  | 30 | –4.93e-15±3.4e-16 | -6.988124e-3 | 457,978 |
| $f_2$ | 10 | 1±0 | 0.9999997 | 457,978 |
|  | 20 | 2±0 | 1.9999993 | 457,978 |
|  | 30 | 3±0 | 2.9999981 | 394,558 |
| $f_3$ | 10 | 1±0 | 0.9999999 | 361,090 |
|  | 20 | 1±3e-18 | 0.9999999 | 294,763 |
|  | 30 | 1±1e-17 | 0.9999999 | 328,243 |
| $f_4$ | 10 | –0.020±0.006 | -0.004429 | 492,372 |
|  | 20 | –0.0026±0.002 | -0.015874 | 361,090 |
|  | 30 | –8.8568e-4±0.001 | -0.002139 | 457,978 |
| $f_5$ | 10 | 0±0 | -1.057361e-4 | 425,640 |
|  | 20 | 0±0 | -1.203252e-3 | 394,558 |
|  | 30 | –5.6843e-16±1e-15 | -9.932735e-5 | 492,372 |
| $f_6$ | 10 | –620.8131±50.4 | -7.753379e-4 | 457,978 |
|  | 20 | –1.3384e+3±68.5 | -7.666976e-4 | 394,558 |
|  | 30 | –2.1395e+3±103.3 | -9.400238e-3 | 294,763 |

\* average best fitness over 20,000,000 evaluations; data reproduced from Table IV in [10] with sign changed because πGASR maximizes $f(x)$ while v-PSO minimizes.

# Appendix 1: Source Code Listing for $\pi$-Fraction Correlation Test Program

```
'Program to test correlation of Pi Fraction coordinates.
'Compiler: PBWIN 10.04.0108 (www.powerbasic.com).

'Last modification: 03-16-2014 ~1501 HRS EST

'(c) Richard A. Formato.  All rights reserved worldwide.
'This source code is FREEWARE.  It may be freely copied,
'distributed and used as long as this copyright notice
'is included without modification and as long as no cost
'or fee is charged to the user, including "bundling" or
'similar fees.  Comments/questions: rf2@ieee.org

#COMPILE EXE
#DIM ALL
%USEMACROS = 1
#INCLUDE "Win32API.inc"
DEFEXT A-Z

GLOBAL PiFractions() AS EXT 'NOTE: Pi Fractions, # fracs, and fraction index are GLOBAL
GLOBAL PiFracIndex&&, IndexIncrement&&, NumPiFractions&&
GLOBAL ScreenWidth&, ScreenHeight&, Quote$

DECLARE FUNCTION RandomInteger&&(n&&,m&&) 'returns random integer I, n=< I < m.
DECLARE FUNCTION RandomNum(a,b)            'returns random number X,  a=< X < b.
DECLARE SUB FillPiFractionArray
DECLARE SUB CreateGNUplotINIfile(PlotWindowULC_X&&,PlotWindowULC_Y&&,PlotWindowWidth&&,PlotWindowHeight&&)
DECLARE SUB TwoDplot(PlotFileName$,PlotTitle$,xCoord$,yCoord$,XaxisLabel$,YaxisLabel$,_
                    LogXaxis$,LogYaxis$,xMin$,xMax$,yMin$,yMax$,xTics$,yTics$,GnuPlotEXE$,LineType$,Annotation$)

'----------------------------------------------------------------------------------------------------------------
FUNCTION PBMAIN () AS LONG
LOCAL Np&&, Nd&&, p&&, i&&, FirstDimNum&&, SecondDimNum&&, N&&, s1$, s2$, s1s2$, DefaultS1S2$, YorN&
LOCAL A$, B$, PlotTitle$, xMin$, xMax$, yMin$, yMax$, NumDims$, OverWritePiFracs$
LOCAL R() AS EXT

    RANDOMIZE TIMER

    DESKTOP GET SIZE TO ScreenWidth&, ScreenHeight&  'get screen size (global variables)
    IF DIR$("wgnuplot.exe") = "" THEN
        MSGBOX("WARNING!  'wgnuplot.exe' not found.  Run terminated.") : EXIT FUNCTION
    END IF
    Quote$ = CHR$(34)

    REDIM PiFractions(1 TO 1)  'NOTE: Sub FillPiFractionArray redims the array
    CALL  FillPiFractionArray  'to the actual number of available fractions
    PiFracIndex&& = NumPiFractions&&*TIMER/86200##  'Sets a variable starting index computed from when the program is started
                                                    'so that each run uses a different set of fractions.  If the same set is
                                                    'desired on succesive runs, set PiFracIndex& = constant of choice.
    OverWritePiFracs$ ="NO"
    YorN& = MSGBOX("Overwrite Pi Fraction array with"+$CRLF+_
                    "pseudorandom (PR) numbers generated by"+$CRLF+_
                    "the compiler using TIMER seed?"+$CRLF+_
                    "Compares Pi Fraction correlation"+$CRLF+_
                    "plots to pseudorandom sequences."+$CRLF,%MB_YESNO+%MB_DEFBUTTON2,"USE PSEUDORANDOM #s INSTEAD")
    A$ = "Pi Fraction index initialized using computer's clock"
    IF YorN& = %IDYES THEN
        OverWritePiFracs$ = "YES"
        A$ = "Pseudorandom array index initialized using computer's clock"
        FOR i&& = 1 TO NumPiFractions&&
            PiFractions(i&&) = RND
        NEXT i&&
    END IF

    YorN& = MSGBOX("Use the computer clock to initialize"+$CRLF+_
                    "the Pi Fraction/PR index? 'Yes' randomizes"+$CRLF+_
                    "the index using the time at which this"+$CRLF+_
                    "program was started.  'No' sets the"+$CRLF+_
                    "initial index to 1, and every run with"+$CRLF+_
                    "this option will use the same set of"+$CRLF+_
                    "Pi Fractions.",%MB_YESNO,"STARTING PI FRACTION/PR INDEX")
    IF YorN& = %IDNO THEN
        PiFracIndex&& = 1 : A$ = "Pi Fraction index initialized to 1" 'A$ = "Pi Fraction/PR index initialized to 1"
    END IF

    s1$ = "-999" : DefaultS1S2$ = "2"
    DO UNTIL VAL(s1$) >= 1 AND VAL(s1$) =< NumPiFractions&&
        s1$ = INPUTBOX$("Pi Fraction/PR Index Increment"+$CRLF+$CRLF+_
                        "[1-"+REMOVE$(STR$(NumPiFractions&&),ANY" ")+"]","INDEX INCREMENT",DefaultS1S2$)
    LOOP
    IndexIncrement&& = VAL(s1$)
    A$ = "["+A$+" with index increment of "+REMOVE$(STR$(IndexIncrement&&),ANY" ")+"]"

    s1$ = "-999" : s2$ = "-999" : DefaultS1S2$ = "30 1000"
    DO UNTIL VAL(s1$) >= 1 AND VAL(s1$) =< 100 AND VAL(s2$) >= 1 AND VAL(s2$) =< 20000
        s1s2$ = INPUTBOX$("# Dimensions and # Data"+$CRLF+"Points to Plot"+$CRLF+_
                            "(separated by a space) "+$CRLF+$CRLF+_
                            "[1-100, 1-20000]","Nd & Np",DefaultS1S2$)
        s1$    = PARSE$(s1s2$,ANY" ./:-_",1)
        s2$    = PARSE$(s1s2$,ANY" ./:-_",2)
    LOOP
    Nd&&    = VAL(s1$) : Np&& = VAL(s2$)
    NumDims$ = REMOVE$(STR$(Nd&&),ANY" ")

    REDIM R(1 TO Np&&, 1 TO Nd&&)
    FOR p&& = 1 TO Np&&
        FOR i&& = 1 TO Nd&&
            IF PiFracIndex&& > NumPiFractions&& THEN PiFracIndex&& = 1
            R(p&&,i&&)          = RandomNum(0##,1##)
        NEXT i&&
    NEXT p&&

    s1$ = "-999" : s2$ = "-999" : DefaultS1S2$ = REMOVE$(STR$(MAX(1,Nd&&-1)),ANY" ")+" "+REMOVE$(STR$(Nd&&),ANY" ")
    DO UNTIL VAL(s1$) >= 1 AND VAL(s1$) =< Nd&& AND VAL(s2$) >= 1 AND VAL(s2$) =< Nd&&
        s1s2$ = INPUTBOX$("1st and 2nd Dimensions to"+$CRLF+_
                            "Plot (separated by a space)"+$CRLF+$CRLF+_
                            "[1-"+NumDims$+", 1-"+NumDims$+"]","PLOTTED COORDINATES",DefaultS1S2$)
        s1$    = PARSE$(s1s2$,ANY" ./:-_",1)
        s2$    = PARSE$(s1s2$,ANY" ./:-_",2)
    LOOP
    FirstDimNum&& = VAL(s1$) : SecondDimNum&& = VAL(s2$)
'----------------------------------------------------------------------------------------------------------------
    B$ = "Pi FRACTION "
    IF OverWritePiFracs$ = "YES" THEN B$ = "PSEUDORANDOM "'change plot title
    PlotTitle$    = B$+"POINTS IN "+REMOVE$(STR$(Nd&&),ANY" ")+" DIMENSIONS, "+_
                    REMOVE$(STR$(Np&&),ANY" ")+" POINTS.\n"+_
                    "Plot of dimensions "+REMOVE$(STR$(FirstDimNum&&),ANY" ")+_
                    " and "+REMOVE$(STR$(SecondDimNum&&),ANY" ")+".\n"+A$
'----------------------------------------------------------------------------------------------------------------
    N&& = FREEFILE
    OPEN "PiFracTst.DAT" FOR OUTPUT AS #N&&
        FOR p&& = 1 TO Np&&
```

```
            PRINT #N&&, USING$("#.#####   #.#####",R(p&&,FirstDimNum&&),R(p&&,SecondDimNum&&))
        NEXT p&&
    CLOSE #N&&
    MSGBOX("Plot data file 'PiFracTst.DAT' has been created.")

    CALL CreateGNUplotINIfile(350,20,1000,1000)
    CALL TwoDplot("PiFracTst.DAT",PlotTitle$,"0.6","0.7","x"+REMOVE$(STR$(FirstDimNum&&),ANY" ")+_
                 "\n\n","\nx"+REMOVE$(STR$(SecondDimNum&&),ANY" ")+"","NO","NO",_
                 xMin$,xMax$,yMin$,yMax$,"5","5","wgnuplot.exe"," pointsize 1 linewidth 2","")
END FUNCTION 'PBMAIN()
'----------------------
FUNCTION RandomNum(a,b)   'returns random number X, a=< X < b.
'<<<<< note: NumPiFractions&&, PiFractions(), PiFracIndex&& are GLOBAL >>>>>
    IF a > b THEN SWAP a, b
    RandomNum     = a + (b-a)*PiFractions(PiFracIndex&&)
    PiFracIndex&& = PiFracIndex&& + IndexIncrement&&
    IF PiFracIndex&& > NumPiFractions&& THEN PiFracIndex&& = 1
END FUNCTION 'RandomNum()
'----------------------
FUNCTION RandomInteger&&(n&&,m&&) 'returns random integer I, n=< I < m.
    IF n&& > m&& THEN SWAP n&&, m&&
    RandomInteger&&  = n&& + (m&&-n&&)*PiFractions(PiFracIndex&&)
    PiFracIndex&&    = PiFracIndex&& + IndexIncrement&&
    IF PiFracIndex&& > NumPiFractions&& THEN PiFracIndex&& = 1
END FUNCTION 'RandomInteger&&()
'---------------------------
SUB FillPiFractionArray
LOCAL N&&, idx&&, Dum$
    IF DIR$("PiFracSUB_215830.DAT") = "" THEN MSGBOX("ERROR!  Pi Frac data file not found!")
    N&& = FREEFILE : OPEN "PiFracSUB_215830.DAT" FOR INPUT AS #N&&
        IF(LOF(N&&)<>2589968) THEN
            MSGBOX("Data file is corrupted!")
        ELSE
            MSGBOX("Data file is "+STR$(LOF(N&&))+" bytes.")
        END IF
        INPUT #N&&, NumPiFractions&&
        REDIM PiFractions(1 TO NumPiFractions&&) 'NOTE FRACTIONS NUMBERED 1 TO #FRACS INSTEAD OF STARTING AT ZERO!
        FOR idx&& = 1 TO NumPiFractions&&
            INPUT #N&&, Dum$
            IF idx&& = 1      AND Dum$ <> "0.141592##" THEN MSGBOX("Error in Pi Frac #1!")
            IF idx&& = 215830 AND Dum$ <> "0.203267##" THEN MSGBOX("Error in Pi Frac #215830!")
            PiFractions(idx&&) = VAL(Dum$)
        NEXT idx&&
    CLOSE #N&&
    MSGBOX("# Pi Fractions is "+STR$(NumPiFractions&&)+".")
END SUB
'------
SUB TwoDplot(PlotFileName$,PlotTitle$,xCoord$,yCoord$,XaxisLabel$,YaxisLabel$, _
             LogXaxis$,LogYaxis$,xMin$,xMax$,yMin$,yMax$,xTics$,yTics$,GnuPlotEXE$,LineType$,Annotation$)
LOCAL N&&, ProcessID???
    N&& = FREEFILE
    OPEN "cmd2d.gp" FOR OUTPUT AS #N&&
        IF LogXaxis$ = "YES" AND LogYaxis$ = "NO"  THEN PRINT #N&&, "set logscale x"
        IF LogXaxis$ = "NO"  AND LogYaxis$ = "YES" THEN PRINT #N&&, "set logscale y"
        IF LogXaxis$ = "YES" AND LogYaxis$ = "YES" THEN PRINT #N&&, "set logscale xy"
        IF xMin$ <> "" AND xMax$ <> "" THEN  PRINT #N&&, "set xrange ["+xMin$+":"+xMax$+"]"
        IF yMin$ <> "" AND yMax$ <> "" THEN  PRINT #N&&, "set yrange ["+yMin$+":"+yMax$+"]"
        PRINT #N&&, "set label "       + Quote$ + Annotation$ + Quote$ + " at graph " + xCoord$ + "," + yCoord$
        PRINT #N&&, "set grid xtics " + xTics$
        PRINT #N&&, "set grid ytics " + yTics$
        PRINT #N&&, "set grid mxtics"
        PRINT #N&&, "set grid mytics"
        PRINT #N&&, "show grid"
        PRINT #N&&, "set title "   + Quote$+PlotTitle$+Quote$
        PRINT #N&&, "set xlabel " + Quote$+XaxisLabel$+Quote$
        PRINT #N&&, "set ylabel " + Quote$+YaxisLabel$+Quote$
        PRINT #N&&, "plot "+Quote$+PlotFileName$+Quote$+" notitle"+LineType$
    CLOSE #N&&
    ProcessID??? = SHELL(GnuPlotEXE$+" cmd2d.gp -")' : CALL Delay(0.5##)
END SUB 'TwoDplot()
'-------------------
SUB CreateGNUplotINIfile(PlotWindowULC_X&&,PlotWindowULC_Y&&,PlotWindowWidth&&,PlotWindowHeight&&)
LOCAL N&&, WinPath$, A$, B$, WindowsDirectory$
    WinPath$ = UCASE$(ENVIRON$("Path"))'DIR$("C:\WINDOWS",23)
    DO
        B$ = A$
        A$ = EXTRACT$(WinPath$,";")
        WinPath$ = REMOVE$(WinPath$,A$+";")
        IF RIGHT$(A$,7) = "WINDOWS" OR A$ = B$ THEN EXIT LOOP
        IF RIGHT$(A$,5) = "WINNT"   OR A$ = B$ THEN EXIT LOOP
    LOOP
    WindowsDirectory$ = A$
    N&& = FREEFILE
'   ---------- WGNUPLOT.INPUT FILE ----------
    OPEN WindowsDirectory$+"\wgnuplot.ini" FOR OUTPUT AS #N&&
        PRINT #N&&,"[WGNUPLOT]"
        PRINT #N&&,"TextOrigin=0 0"
        PRINT #N&&,"TextSize=640 150"
        PRINT #N&&,"TextFont=Terminal,9"
        PRINT #N&&,"GraphOrigin="+REMOVE$(STR$(PlotWindowULC_X&&),ANY" ")+" "+REMOVE$(STR$(PlotWindowULC_Y&&),ANY" ")
        PRINT #N&&,"GraphSize="  +REMOVE$(STR$(PlotWindowWidth&&),ANY" ")+" "+REMOVE$(STR$(PlotWindowHeight&&),ANY" ")
        PRINT #N&&,"GraphFont=Arial,10"
        PRINT #N&&,"GraphColor=1"
        PRINT #N&&,"GraphToTop=1"
        PRINT #N&&,"GraphBackground=255 255 255"
        PRINT #N&&,"Border=0 0 0 0"
        PRINT #N&&,"Axis=192 192 192 2 2"
        PRINT #N&&,"Line1=0 0 255 0 0"
        PRINT #N&&,"Line2=0 255 0 0 1"
        PRINT #N&&,"Line3=255 0 0 0 2"
        PRINT #N&&,"Line4=255 0 255 0 3"
        PRINT #N&&,"Line5=0 0 128 0 4"
    CLOSE #N&&
END SUB 'CreateGNUplotINIfile()
'--------------------------------------------- END  PROGRAM ---------------------------------------------
```

# Appendix 2: Source Code Listing for Algorithm $\pi$GASR

```
'Program GASR_PI_FRAC_01-13-2014.BAS.
'Stand-alone Genetic Algorithm with Sibling
'Rivalry for use in Hybrid CFO-GASR Algorithm
'utilizing Pi Fractions as pseudorandom variables.

'Compiled with Power Basic/Windows Compiler 10.04.0108 (www.PowerBasic.com).

'LAST MOD 01-13-2014 ~0831 HRS EST
'------------------------------------------------------------------------
'This program was used to generate data for the
'arXiv paper posted 13 January 2014.

'(c) 2014 Richard A. Formato

'This material is FREEWARE.  It may be freely copied and distributed without
'limitation as long as there is no charge of any kind, including "tie-in",
'"bundling," or similar fees or charges.

#COMPILE EXE
#DIM ALL
%USEMACROS = 1
#INCLUDE "win32API.inc"
DEFEXT A-Z
'---------------------
GLOBAL NumPiFractions&&

GLOBAL PiFractions() AS EXT

GLOBAL RunDataFile$, CheckEarlyTerm$

GLOBAL EulerConst, Pi, Pi2, Pi4, TwoPi, FourPi, FivePi, e, Root2 AS EXT

GLOBAL Alphabet$, Digits$, RunID$, Quote$, SpecialCharacters$

GLOBAL Mu0, Eps0, c, eta0 AS EXT

GLOBAL RadToDeg, DegToRad, Feet2Meters, Meters2Feet, Inches2Meters, Meters2Inches AS EXT

GLOBAL Miles2Meters, Meters2Miles, NautMi2Meters, Meters2NautMi AS EXT
'----------------------------------------------------------------=-----------------------------------
DECLARE SUB Get_Pi_Fractions

DECLARE FUNCTION RandomIntegerPiFrac%(n%,m%,k&&)

DECLARE FUNCTION RandomNumPiFrac(a,b,k&&)

DECLARE SUB RunGASR(BestFitnessAllRuns,GASRseedChromo(),FunctionName$,Nd%,XiMin(),XiMax(),_
         TotalEvalsThisGroup&&,GAgroupNum%,NumGAgroups%,LastGenNum&)

DECLARE FUNCTION ObjectiveFunction(R(),Nd%,p%,j&,FunctionName$)

DECLARE FUNCTION Colville(R(),Nd%,p%,j&)      'Colville (4-D)

DECLARE FUNCTION Griewank(R(),Nd%,p%,j&)      'Griewank (n-D)

DECLARE FUNCTION Ackley(R(),Nd%,p%,j&)        'Ackley

DECLARE FUNCTION Exponential(R(),Nd%,p%,j&)   'Exponential

DECLARE FUNCTION CosineMix(R(),Nd%,p%,j&)     'Cosine Mix

DECLARE FUNCTION ParrottF4(R(),Nd%,p%,j&)     'Parrott F4 (1D)

DECLARE FUNCTION Schwefel(R(),Nd%,p%,j&)      'Schwefel Problem 2.26 (nD)

DECLARE FUNCTION Rastrigin(R(),Nd%,p%,j&)     'Rastrigin (nD)

DECLARE FUNCTION SGO(R(),Nd%,p%,j&)           'SGO (Space Gravitational Optimization)

DECLARE FUNCTION GP(R(),Nd%,p%,j&)            'Goldstein-Price

DECLARE SUB GA(FunctionName$,Np%,Nd%,NumGenerations&,XiMin(),XiMax(),Chromos(),Fitness(),BestChromoThisRun%,_
         BestGenerationThisRun&,BestFitnessThisRun,NevalThisRun&&,BestFitnessVsGenThisRun(),LastGenNum&)

DECLARE SUB MutateChromo(XiMin(),XiMax(),Chromos(),ChromoNum%,Nd%,GenNum&,NumGenerations&,Alpha,w)

DECLARE SUB MathematicalConstants

DECLARE SUB AlphabetAndDigits

DECLARE SUB SpecialSymbols

DECLARE SUB EMconstants

DECLARE SUB ConversionFactors
'=====================================================================================================
FUNCTION PBMAIN () AS LONG

LOCAL i%, Nd%, m%, TotalEvalsThisGroup&&, TotalEvalsOverAllGroups&&, NumGAgroups%

LOCAL BestFitnessAllRuns, GASRseedChromo(), XiMin(), XiMax() AS EXT

LOCAL FunctionName$, A$, Coords$, LastGenNum&
'---------- Globals -----------
CALL AlphabetAndDigits
CALL MathematicalConstants
CALL SpecialSymbols
CALL EMconstants
CALL ConversionFactors
CALL Get_Pi_Fractions 'fills GLOBAL Pi fraction array

'--------------------------- Test Function -------------------------------

    CheckEarlyTerm$ = "YES" '"NO"

    FunctionName$   = "Schwefel" '"Rastrigin" '"Griewank" '"Expon" '"CosMix" '"Ackley"

    SELECT CASE FunctionName$
        CASE "SGO"       : Nd% = 2
        CASE "GP"        : Nd% = 2
        CASE "Rastrigin" : Nd% = 30
        CASE "Schwefel"  : Nd% = 30
        CASE "ParrottF4" : Nd% = 1
        CASE "Expon"     : Nd% = 30
        CASE "Ackley"    : Nd% = 30
        CASE "CosMix"    : Nd% = 30
```

```
            CASE "Griewank" : Nd% = 30
            CASE "Colville" : Nd% = 4
            CASE ELSE : MSGBOX("ERROR in FunctionName$ !!") : EXIT FUNCTION
        END SELECT

        REDIM XiMin(1 TO Nd%), XiMax(1 TO Nd%), GASRseedChromo(1 TO Nd%)

        SELECT CASE FunctionName$
            CASE "SGO"      : FOR i% = 1 TO Nd% : XiMin(i%) =  -50## : XiMax(i%) =   50## : NEXT i% 'DS for SGO
            CASE "GP"       : FOR i% = 1 TO Nd% : XiMin(i%) = -100## : XiMax(i%) =  100## : NEXT i% 'DS for Goldstein-Price
            CASE "Rastrigin": FOR i% = 1 TO Nd% : XiMin(i%) =  -10## : XiMax(i%) =   10## : NEXT i% 'Rastrigin boundaries from Li A&P paper (note: larger than
the usual +/-5.12)
            CASE "Schwefel" : FOR i% = 1 TO Nd% : XiMin(i%) = -500## : XiMax(i%) =  500## : NEXT i% 'DS for Schwefel 2.26
            CASE "ParrottF4": FOR i% = 1 TO Nd% : XiMin(i%) =    0## : XiMax(i%) =    1## : NEXT i% 'DS for ParrottF4
            CASE "Expon"    : FOR i% = 1 TO Nd% : XiMin(i%) =   -1## : XiMax(i%) =    1## : NEXT i% 'DS for Exponential
            CASE "Ackley"   : FOR i% = 1 TO Nd% : XiMin(i%) =  -30## : XiMax(i%) =   30## : NEXT i% 'DS for Ackley
            CASE "CosMix"   : FOR i% = 1 TO Nd% : XiMin(i%) =   -1## : XiMax(i%) =    1## : NEXT i% 'DS for Cosine Mix
            CASE "Griewank" : FOR i% = 1 TO Nd% : XiMin(i%) = -600## : XiMax(i%) =  600## : NEXT i% 'DS for Griewank
            CASE "Colville" : FOR i% = 1 TO Nd% : XiMin(i%) =  -10## : XiMax(i%) =   10## : NEXT i% 'DS for Colville
        END SELECT
'------------
        A$                   = ""
        NumGAgroups%         = 1
        TotalEvalsOverAllGroups&& = 0
        FOR m% = 1 TO NumGAgroups% 'NOTE: Option of executing NumGAgroups%>1 groups of independent GA runs as consistency check
                                   '     if compiler built-in RND function is used for randomization instead of Pi fractions.
            CALL RunGASR(BestFitnessAllRuns,GASRseedChromo(),FunctionName$,Nd%,XiMin(),XiMax(),TotalEvalsThisGroup&&,m%,_
                         NumGAgroups%,LastGenNum&)
            TotalEvalsOverAllGroups&& = TotalEvalsOverAllGroups&& + TotalEvalsThisGroup&&
            Coords$ = "" : FOR i% = 1 TO Nd% : Coords$ = Coords$+SPACE$(30)+"i ="+STR$(i%)+_
                          ": "+STR$(ROUND(GASRseedChromo(i%),6)) + $CRLF : NEXT i%
            A$ = A$+"GA group #"+STR$(m%)+": Best Fitness = "+STR$(BestFitnessAllRuns,12)+" @  coords:"+$CRLF+$CRLF+Coords$
        NEXT m%

        MSGBOX("in PBMAIN, "+REMOVE$(STR$(Nd%),ANY "  ")+"D "+FunctionName$+_
                    ": Tot Evals = "+STR$(TotalEvalsOverAllGroups&&)+" / "+STR$(LastGenNum&+1)+" Gens"+$CRLF+$CRLF+_
                    A$+$CRLF+"Output File(s):   "+FunctionName$+"*.DAT")
END FUNCTION 'PBMAIN()
'--------------------

SUB Get_Pi_Fractions

LOCAL p&&, i&&, K&&, m&&, N&&, LineCount&&, idx&&, Lyne$, PiDigits$(), DataFileName$, A$

LOCAL LastDigit&&, ThisDigit&&, NumBins&&, BinNum&&, TotalCount&&, PDF&&(), PDFtotalPoints&&

LOCAL BinMin, BinMax, CDF(), AvgPtsPerBin AS EXT

LOCAL PiFracAvg AS EXT

'NOTE: The user must replace this code to read an external Pi fraction data file that
'      probably has a different name and format than file "BBP_Pi_160K.DAT".

    DataFileName$ = "BBP_Pi_160K.DAT"
    IF DIR$(DataFileName$) = "" THEN
        MSGBOX("ERROR! Data file "+DataFileName$+" not found.  Run terminated.") : END
    END IF

    LineCount&& = 0 : NumPiFractions&& = 0
    N&& = FREEFILE  : OPEN DataFileName$ FOR INPUT AS #N&&
    K&& = FREEFILE  : OPEN "TEMP" FOR OUTPUT AS #K&&
        WHILE NOT EOF(N&&)
            INPUT #N&&, Lyne$ : Linecount&& = Linecount&& + 1
            IF INSTR(Lyne$,"fraction") > 0 THEN
                PRINT #K&&, Lyne$ : NumPiFractions&& = NumPiFractions&& + 1
            END IF
        WEND
    CLOSE K&& : CLOSE #N&&

    REDIM PiDigits$(1 TO NumPiFractions&&), PiFractions(1 TO NumPiFractions&&)

    N&& = FREEFILE : OPEN "TEMP" FOR INPUT AS #N&&
        FOR m&& = 1 TO NumPiFractions&& : INPUT #N&&, Lyne$ : PiDigits$(m&&) = Lyne$ : NEXT m&&
    CLOSE #N&&
    KILL "TEMP"

    PiFracAvg = 0##
    FOR m&& = 1 TO NumPiFractions&&
        PiFractions(m&&) = VAL(REMOVE$(PiDigits$(m&&),ANY Alphabet$+" ="))
        PiFracAvg        = PiFracAvg + PiFractions(m&&)
    NEXT m&&
    PiFracAvg = PiFracAvg/NumPiFractions&&

'   ---------- PDF/CDF ----------

    NumBins&& = 1000 'Compute PDF
    REDIM PDF&&(1 TO NumBins&&),CDF(1 TO NumBins&&)
    FOR BinNum&& = 1 TO NumBins&&
        BinMin = (BinNum&&-1)/NumBins&& : BinMax = BinNum&&/NumBins&& 'Pi fractions are in [0,1]
        FOR m&& = 1 TO NumPiFractions&&
            IF PiFractions(m&&) >= BinMin AND PiFractions(m&&) < BinMax THEN PDF&&(BinNum&&) = PDF&&(BinNum&&) + 1
        NEXT m&&
    NEXT BinNum&&

    PDFtotalPoints&& = 0
    FOR BinNum&& = 1 TO NumBins&& : PDFtotalPoints&& = PDFtotalPoints&& + PDF&&(BinNum&&) : NEXT BinNum&&
    IF PDFtotalPoints&& <> NumPiFractions&& THEN
        MSGBOX("ERROR! Tot PDF points ="+STR$(PDFtotalPoints&&)+" <> # Pi fractions ("+REMOVE$(STR$(NumPiFractions&&),ANY " ")+").")
        END
    END IF
    AvgPtsPerBin = PDFtotalPoints&&/NumBins&&

    CDF(1) = PDF&&(1)/NumPiFractions&&
    FOR BinNum&& = 2 TO NumBins&&   'Compute CDF
        CDF(BinNum&&) = CDF(BinNum&&-1) + PDF&&(BinNum&&)/NumPiFractions&&
    NEXT BinNum&&

    TotalCount&& = 0
    N&& = FREEFILE : OPEN "Pi_Frac_PDF&CDF.DAT" FOR OUTPUT AS #N&&
        PRINT #N&&,"Pi Fraction Statistical Data"+$CRLF+"Created "+DATE$+" "+TIME$+$CRLF
        PRINT #N&&,"#fractions:"+STR$(NumPiFractions&&)+$CRLF+"Mean value: 0"+REMOVE$(STR$(PiFracAvg),ANY " ")+$CRLF
        PRINT #N&&," Norm Bin#     PDF        CDF"
        A$ =        "#.########  #.####    #.########"
        PRINT #N&&, USING$(A$,0,1,0)
        FOR BinNum&& = 1 TO NumBins&&
            TotalCount&& = TotalCount&& + PDF&&(BinNum&&)
            PRINT #N&&, USING$(A$,BinNum&&/NumBins&&,PDF&&(BinNum&&)/AvgPtsPerBin,CDF(BinNum&&))
        NEXT BinNum&&
    CLOSE #N&&
    MSGBOX("File 'Pi_Frac_PDF&CDF.DAT' created."   +$CRLF+_
           "Total Data Points ="+STR$(TotalCount&&)+$CRLF+_
           "Average Value = "   +STR$(PiFracAvg))

END SUB 'Get_Pi_Fractions

'-----------------------
```

```
SUB RunGASR(BestFitnessAllRuns,GASRseedChromo(),FunctionName$,Nd%,XiMin(),XiMax(),TotalEvalsThisGroup&&,GAgroupNum%,NumGAgroups%,LastGenNum&)

'Creates a Group of Independent GA runs and returns Best Fitness
'and corresponding Chromosome across all runs in all groups.

LOCAL Chromos(), Fitness(), BestFitnessThisRun, BestFitnessVsGenThisRun() AS EXT

LOCAL i%, N%, BestChromoThisRun%, BestGenerationThisRun&, NevalThisRun&&, A$

LOCAL RunNumber%, GenNum&, NumGenerations&, NumChromos%, NumIndependentGArunsPerGroup%

'-------------------------------------------------
NumIndependentGArunsPerGroup% =    1
NumGenerations&               =  100
NumChromos%                   = 2500  'MUST BE EVEN #
'-------------------------------------------------

BestFitnessAllRuns = -1E4200 : TotalEvalsThisGroup&& = 0

REDIM Chromos(1 TO NumChromos%, 1 TO Nd%, 0 TO NumGenerations&),_
      Fitness(1 TO NumChromos%, 0 TO NumGenerations&),_
      BestFitnessVsGenThisRun(0 TO NumGenerations&)

'NOTE: GA generations are numbered starting with ZERO for consistency with the iteration
'      numbering in CFO (anticpating that GA might be more tightly integrated into CFO).

REDIM GASRseedChromo(1 TO Nd%)

FOR RunNumber% = 1 TO NumIndependentGArunsPerGroup%

    CALL GA(FunctionName$,NumChromos%,Nd%,NumGenerations&,XiMin(),XiMax(),Chromos(),Fitness(),_
            BestChromoThisRun%,BestGenerationThisRun&,BestFitnessThisRun,NevalThisRun&&,_
            BestFitnessVsGenThisRun(),LastGenNum&)

    TotalEvalsThisGroup&& = TotalEvalsThisGroup&& + NevalThisRun&&

    IF BestFitnessThisRun >= BestFitnessAllRuns THEN
        BestFitnessAllRuns = BestFitnessThisRun
        FOR i% = 1 TO Nd% : GASRseedChromo(i%) = Chromos(BestChromoThisRun%,i%,BestGenerationThisRun&) : NEXT i%
    END IF

    RunDataFile$ = FunctionName$+"_"+REMOVE$(STR$(Nd%),ANY " ")+"D_GR#"+REMOVE$(STR$(GAgroupNum%),ANY " ")+_
                                     "_Run#"+REMOVE$(STR$(RunNumber%),ANY " ")+".DAT"
    N% = FREEFILE
    OPEN RunDataFile$ FOR OUTPUT AS #N%
        A$ = REMOVE$(STR$(Nd%),ANY " ")+"D "+FunctionName$
        PRINT# N%, A$+$CRLF+STRING$(LEN(A$),"-")                   +$CRLF+_
                   "Run ID:  " + RunID$                             +$CRLF+_
                   "# Groups: "+STR$(NumGAgroups%)                  +$CRLF+_
                   "# GA runs/group: "+STR$(NumIndependentGArunsPerGroup%) +$CRLF+_
                   "# Generations:   "  +STR$(NumGenerations&)      +$CRLF+_
                   "# Chromos:       "  +STR$(NumChromos%)          +$CRLF+_
                   "# Eval this run: "  +STR$(NevalThisRun&&)       +$CRLF+_
                   "# Gens req'd:    "  +STR$(LastGenNum&+1)        +$CRLF+$CRLF+_
                   "     Gen #         Best Fitness"                +$CRLF+_
                   "    --------      ------------"
        A$=      "    ######      ######.#####"
        FOR GenNum& = 0 TO LastGenNum& : PRINT #N%, USING$(A$,GenNum&,BestFitnessVsGenThisRun(GenNum&)) : NEXT GenNum&
    CLOSE #N%

NEXT RunNumber%

END SUB 'RunGASR()

'--------------------------
SUB GA(FunctionName$,NumChromos%,Nd%,NumGenerations&,XiMin(),XiMax(),Chromos(),Fitness(),_
       BestChromoThisRun%,BestGenerationThisRun&,BestFitnessThisRun,NevalThisRun&&,BestFitnessVsGenThisRun(),LastGenNum&)

'GENETIC ALGORITHM WITH SIBLING RIVALRY (Li et al.)
'-------------------------------------------------
'GASR Reference: Li W. T., Shi X. W., Hei Y. Q., Liu S. F., and Zhu J., "A Hybrid Optimization Algorithm and Its Application for
'Conformal Array Pattern Synthesis," IEEE Trans. Ant. & Prop., vol.58, no.10, October 2010, pp.3401-3406.

LOCAL ChromoNum%, i%, GenNum& 'chromo # (a/k/a probe,individual,agent); gene # (a/k/a decision variable, dimension);
                               'generation # (a/k/a time step, iteration) [terminology in other metaheuristics].

LOCAL CrossOverProbability, MutationProbability, w, Alpha, MAXI, MINI, SUM2, BestPreviousFitness, Tol AS EXT

LOCAL b1(), b2(), b3(), b4(), ChildChromoFitness() AS EXT 'child chromos

LOCAL ChildChromoNumbersOrderedByFitness%(), NumberOfBestPreviousChromo%, NumberOfSecondBestPreviousChromo%, ElitistChromoNumber1%

LOCAL L%, M%, s%, t%, k&& 'k&& is Pi Fraction index

LOCAL A$

'------------------------------------------- Initialize Matrices -------------------------------------------

REDIM Chromos(1 TO NumChromos%, 1 TO Nd%, 0 TO NumGenerations&), Fitness(1 TO NumChromos%, 0 TO NumGenerations&)

REDIM b1(1 TO NumChromos%, 1 TO Nd%, 0 TO NumGenerations&), b2(1 TO NumChromos%, 1 TO Nd%, 0 TO NumGenerations&),_
      b3(1 TO NumChromos%, 1 TO Nd%, 0 TO NumGenerations&), b4(1 TO NumChromos%, 1 TO Nd%, 0 TO NumGenerations&)

REDIM BestFitnessVsGenThisRun(0 TO NumGenerations&)

REDIM ChildChromoNumbersOrderedByFitness%(1 TO 4)

LastGenNum& = NumGenerations&

FOR L% = 1 TO 4 : ChildChromoNumbersOrderedByFitness%(L%) = L% : NEXT L%

'-------------------------------------- Run GA --------------------------------------

'NOTE: THE FOLLOWING PARAMETERS ARE SET EMPIRICALLY BASED ON GETTING 'GOOD' RESULTS...
CrossOverProbability  = 0.8##
MutationProbability   = 0.02##
w                     = 0.5##
Alpha                 = 2##
Tol                   = 0.000001##  'for early termination

BestPreviousFitness = -1E4200 : BestFitnessThisRun = -1E4200 : NevalThisRun&& = 0

'STEP (A): CREATE RANDOM INITIAL POPULATION (GENERATION #0) AND COMPUTE INITIAL FITNESSES

    FOR ChromoNum% = 1 TO NumChromos%
        FOR i% = 1 TO Nd%
'IMPORTANT NOTE: If more than one run/group is made and/or more than one group is used, then the run and
'                group numbers should be included in the Pi Fraction index to avoid using the same fractions
'                in successive runs.  Note too that any scheme for varying the index should work well because
'                the Pi Fractions are uniformly distributed in (0,1), but that the scheme should avoid over-
'                lapping too many fractions (if any) in order to avoid creating multiple copies of a single
'                point in the decision space.
            k&& = ChromoNum%*i%
            Chromos(ChromoNum%,i%,0) = RandomNumPiFrac(XiMin(i%),XiMax(i%),k&&)
        NEXT i%
    NEXT ChromoNum%
```

```
       FOR ChromoNum% = 1 TO NumChromos%
             Fitness(ChromoNum%,0)      = ObjectiveFunction(Chromos(),Nd%,ChromoNum%,0,FunctionName$)
             INCR NevalThisRun&&
             IF Fitness(ChromoNum%,0) >= BestPreviousFitness THEN
                  NumberOfBestPreviousChromo% = ChromoNum%
                  BestPreviousFitness         = Fitness(ChromoNum%,0)
                  BestFitnessVsGenThisRun(0)  = Fitness(ChromoNum%,0)
             END IF
             IF Fitness(ChromoNum%,0)       >= BestFitnessThisRun THEN
                  BestChromoThisRun%     = ChromoNum%
                  BestGenerationThisRun& = 0
                  BestFitnessThisRun     = Fitness(ChromoNum%,0)
             END IF
       NEXT ChromoNum%

FOR GenNum& = 1 TO NumGenerations& 'Loop over generations

     FOR ChromoNum% = 1 TO NumChromos%-1 STEP 2 'Loop over chromos pairwise in each generation

       k&& = ChromoNum%+GenNum&
       IF RandomNumPiFrac(0##,1##,k&&) =< CrossOverProbability THEN 'Create 2 new chromos this gen from children
                                                                   'whose parents are from previous generation.
  'STEP (B): SELECT TWO DIFFERENT PARENT CHROMOS FROM _PREVIOUS_ GENERATION
       k&& = ChromoNum%
       DO
             s% = RandomIntegerPiFrac%(1,NumChromos%,k&&) : t% = RandomIntegerPiFrac%(1,NumChromos%,k&&+2) 's, t are parent chromo #s
             IF s% <> t% THEN EXIT LOOP
             k&& = k&& + 1
       LOOP

  'STEP (C): CROSSOVER (MATE) PARENTS TO CREATE FOUR CHILD CHROMOS
       FOR i% = 1 TO Nd%
             MAXI = MAX(Chromos(s%,i%,GenNum&-1),Chromos(t%,i%,GenNum&-1))
             MINI = MIN(Chromos(s%,i%,GenNum&-1),Chromos(t%,i%,GenNum&-1))
             SUM2 = (Chromos(s%,i%,GenNum&-1)+Chromos(t%,i%,GenNum&-1))/2##

             b1(ChromoNum%,i%,GenNum&) = (1##-w) * MAXI      + w*SUM2
             b2(ChromoNum%,i%,GenNum&) = (1##-w) * MINI      + w*SUM2
             b3(ChromoNum%,i%,GenNum&) = (1##-w) * XiMax(i%) + w*MAXI
             b4(ChromoNum%,i%,GenNum&) = (1##-w) * XiMin(i%) + w*MINI
       NEXT i%

  ' --------------------------------- Choose Two Best Child Chromos ------------------------------------

       REDIM ChildChromoFitness(1 TO 4)
       ChildChromoFitness(1) = ObjectiveFunction(b1(),Nd%,ChromoNum%,GenNum&,FunctionName$) : INCR NevalThisRun&&
       ChildChromoFitness(2) = ObjectiveFunction(b2(),Nd%,ChromoNum%,GenNum&,FunctionName$) : INCR NevalThisRun&&
       ChildChromoFitness(3) = ObjectiveFunction(b3(),Nd%,ChromoNum%,GenNum&,FunctionName$) : INCR NevalThisRun&&
       ChildChromoFitness(4) = ObjectiveFunction(b4(),Nd%,ChromoNum%,GenNum&,FunctionName$) : INCR NevalThisRun&&

       ARRAY SORT ChildChromoFitness(), TAGARRAY ChildChromoNumbersOrderedByFitness%(), DESCEND
       'Array ChildChromoNumbersOrderedByFitness%() contains chromo numbers, best to worst.
  ' ---------------------------- Copy Best Child Chromos Into New Chromos in This Generation -------------------------------

       SELECT CASE ChildChromoNumbersOrderedByFitness%(1) '1st new chromo this gen, #ChromoNum%
             CASE 1 : FOR i% = 1 TO Nd% : Chromos(ChromoNum%,i%,GenNum&) = b1(ChromoNum%,i%,GenNum&)   : NEXT i% 'best child is b1()
             CASE 2 : FOR i% = 1 TO Nd% : Chromos(ChromoNum%,i%,GenNum&) = b2(ChromoNum%,i%,GenNum&)   : NEXT i% 'best child is b2()
             CASE 3 : FOR i% = 1 TO Nd% : Chromos(ChromoNum%,i%,GenNum&) = b3(ChromoNum%,i%,GenNum&)   : NEXT i% 'best child is b3()
             CASE 4 : FOR i% = 1 TO Nd% : Chromos(ChromoNum%,i%,GenNum&) = b4(ChromoNum%,i%,GenNum&)   : NEXT i% 'best child is b4()
       END SELECT

       SELECT CASE ChildChromoNumbersOrderedByFitness%(2) '2nd new chromo this gen, #ChromoNum%+1
             CASE 1 : FOR i% = 1 TO Nd% : Chromos(ChromoNum%+1,i%,GenNum&) = b1(ChromoNum%,i%,GenNum&) : NEXT i% '2nd best child is b1()
             CASE 2 : FOR i% = 1 TO Nd% : Chromos(ChromoNum%+1,i%,GenNum&) = b2(ChromoNum%,i%,GenNum&) : NEXT i% '2nd best child is b2()
             CASE 3 : FOR i% = 1 TO Nd% : Chromos(ChromoNum%+1,i%,GenNum&) = b3(ChromoNum%,i%,GenNum&) : NEXT i% '2nd best child is b3()
             CASE 4 : FOR i% = 1 TO Nd% : Chromos(ChromoNum%+1,i%,GenNum&) = b4(ChromoNum%,i%,GenNum&) : NEXT i% '2nd best child is b4()
       END SELECT

ELSE 'If RandomNumPiFrac(0##,1##,k&&)>CrossOverProbability, then copy into this
     'gen with no change the two corresponding chromos from previous generation.

       FOR i% = 1 TO Nd%:  Chromos(ChromoNum%,i%,GenNum&)   = Chromos(ChromoNum%,i%,GenNum&-1)   : NEXT i%
       FOR i% = 1 TO Nd%:  Chromos(ChromoNum%+1,i%,GenNum&) = Chromos(ChromoNum%+1,i%,GenNum&-1) : NEXT i%

END IF 'RandomNumPiFrac(0##,1##,k&&) =< CrossOverProbability

  'STEP (D): MUTATE EACH OF THE TWO NEW CHROMOS

       k&& = ChromoNum%+GenNum&
       IF RandomNumPiFrac(0##,1##,k&&) =< MutationProbability THEN 'mutate chromos, otherwise don't.
             CALL MutateChromo(XiMin(),XiMax(),Chromos(),ChromoNum%,Nd%,GenNum&,NumGenerations&,Alpha,w)
             CALL MutateChromo(XiMin(),XiMax(),Chromos(),ChromoNum%+1,Nd%,GenNum&,NumGenerations&,Alpha,w)
       END IF

     NEXT ChromoNum% 'Chromo Loop

  'STEP (E): ELITISM - INSERT BEST PREVIOUS CHROMO RANDOMLY INTO THIS GENERATION (MAYBE INSERT TWO BEST??)
       k&&                 = GenNum&
       ElitistChromoNumber1% = RandomIntegerPiFrac%(1,NumChromos%,k&&)
       FOR i% = 1 TO Nd% : Chromos(ElitistChromoNumber1%,i%,GenNum&) =_ Chromos(NumberOfBestPreviousChromo%,i%,GenNum&-1) : NEXT i%

  'STEP (F): COMPUTE FITNESS FOR ALL CHROMOS IN THIS GENERATION

       FOR ChromoNum% = 1 TO NumChromos%

          Fitness(ChromoNum%,GenNum&) = ObjectiveFunction(Chromos(),Nd%,ChromoNum%,GenNum&,FunctionName$) : INCR NevalThisRun&&

          IF Fitness(ChromoNum%,GenNum&) >= BestPreviousFitness THEN
                  NumberOfBestPreviousChromo% = ChromoNum% : BestPreviousFitness = Fitness(ChromoNum%,GenNum&)
                  BestFitnessVsGenThisRun(GenNum&) = Fitness(ChromoNum%,GenNum&)
                  IF CheckEarlyTerm$ = "YES" AND GenNum& > 20 AND GenNum& MOD 5 = 0 AND _
                     BestFitnessVsGenThisRun(GenNum&)-BestFitnessVsGenThisRun(GenNum&-19) =< Tol THEN 'Early termination test.
                        LastGenNum& = GenNum& : EXIT SUB
                  END IF
          END IF

          IF Fitness(ChromoNum%,GenNum&) >= BestFitnessThisRun THEN
                  BestChromoThisRun%     = ChromoNum%
                  BestGenerationThisRun& = GenNum&
                  BestFitnessThisRun     = Fitness(ChromoNum%,GenNum&)
          END IF

       NEXT ChromoNum%

NEXT GenNum& 'Generation Loop

END SUB 'GA()
'------------
SUB MutateChromo(XiMin(),XiMax(),Chromos(),ChromoNum%,Nd%,GenNum&,NumGenerations&,Alpha,w)

LOCAL K%, m&&

LOCAL Xk, XkU, XkL, XkStar, Mu AS EXT
```

```
    m&&  = ChromoNum%+GenNum&
    K%   = RandomIntegerPiFrac%(1,Nd%,m&&) 'randomly select element index K for mutation

    Xk   = Chromos(ChromoNum%,K%,GenNum&) 'Kth element of chromo BEFORE mutation

    Mu   = 1## -(GenNum&/NumGenerations&)^((1##-GenNum&/NumGenerations&)^Alpha) 'Alpha is the "shape parameter"

    XkU  = MIN(Xk + Mu*(XiMax(K%)-XiMin(K%))/2##,XiMax(K%))

    XkL  = MAX(Xk - Mu*(XiMax(K%)-XiMin(K%))/2##,XiMin(K%))

    XkStar = XkL + w*(XkU-XkL) 'Kth element of chromo AFTER mutation

    Chromos(ChromoNum%,K%,GenNum&) = XkStar 'insert Kth element after mutating it

END SUB 'MutateChromo()
'---------------------
FUNCTION ObjectiveFunction(R(),Nd%,p%,j&,FunctionName$) 'Objective function to be MAXIMIZED is defined here
    SELECT CASE FunctionName$
        CASE "GP"       : ObjectiveFunction = GP(R(),Nd%,p%,j&)          'Goldstein-Price (2D)
        CASE "SGO"      : ObjectiveFunction = SGO(R(),Nd%,p%,j&)         'SGO Function (2D)
        CASE "Rastrigin": ObjectiveFunction = Rastrigin(R(),Nd%,p%,j&)   'Rastrigin (nD)
        CASE "Schwefel" : ObjectiveFunction = Schwefel(R(),Nd%,p%,j&)    'Schwefel 2.26 (nD)
        CASE "ParrottF4": ObjectiveFunction = ParrottF4(R(),Nd%,p%,j&)   'Parrott F4 (1D)
        CASE "Expon"    : ObjectiveFunction = Exponential(R(),Nd%,p%,j&) 'Exponential (30D)
        CASE "Ackley"   : ObjectiveFunction = Ackley(R(),Nd%,p%,j&)      'Ackley (30D)
        CASE "CosMix"   : ObjectiveFunction = CosineMix(R(),Nd%,p%,j&)   'Cosine Mix (30D)
        CASE "Griewank" : ObjectiveFunction = Griewank(R(),Nd%,p%,j&)    'Griewank (30D)
        CASE "Colville" : ObjectiveFunction = Colville(R(),Nd%,p%,j&)    'Colville (4D)
    END SELECT
END FUNCTION 'ObjectiveFunction()
'-----------------------------
FUNCTION ParrottF4(R(),Nd%,p%,j&) 'Parrott F4 (1D)
'MAXIMUM = 1 AT ~0.0796875... WITH ZERO OFFSET (SEEMS TO WORK BEST WITH JUST 3 PROBES, BUT NOT ALLOWED IN THIS VERSION...)
'References: (1) Beasley, D., D. R. Bull, and R. R. Martin, "A Sequential Niche Technique for Multimodal
'Function Optimization," Evol. Comp. (MIT Press), vol. 1, no. 2, 1993, pp. 101-125
'(online at http://citeseer.ist.psu.edu/beasley93sequential.html).
'(2) Parrott, D., and X. Li, "Locating and Tracking Multiple Dynamic Optima by a Particle Swarm
'Model Using Speciation," IEEE Trans. Evol. Computation, vol. 10, no. 4, Aug. 2006, pp. 440-458.

LOCAL Z, x, offset AS EXT
    offset = 0##

    x = R(p%,1,j&)

    Z = EXP(-2##*LOG(2##)*((x-0.08##-offset)/0.854##)^2)*(SIN(5##*Pi*((x-offset)^0.75##-0.05##)))^6
'WARNING! This is a NATURAL LOG, NOT Log10!!!

    ParrottF4 = Z
END FUNCTION 'ParrottF4()
'------------------------
FUNCTION Schwefel(R(),Nd%,p%,j&) 'Schwefel Problem 2.26 (n-D)
'MAXIMUM = 12,569.5 @ [420.8687]^30 (30-D CASE).
'Reference: Yao, X., Liu, Y., and Lin, G., "Evolutionary Programming Made Faster,"
'IEEE Trans. Evolutionary Computation, Vol. 3, No. 2, 82-102, Jul. 1999.

    LOCAL Z, Xi AS EXT
    LOCAL i%
    Z = 0##
    FOR i% = 1 TO Nd%
        Xi = R(p%,i%,j&)
        Z = Z + Xi*SIN(SQR(ABS(Xi)))
    NEXT i%
    Schwefel = Z - 418.9829##*Nd%
END FUNCTION 'SCHWEFEL226()
'------------------------
FUNCTION Rastrigin(R(),Nd%,p%,j&) 'Rastrigin (nD)
'MAXIMUM = ZERO (nD CASE).
'Reference: Yao, X., Liu, Y., and Lin, G., "Evolutionary Programming Made Faster,"
'IEEE Trans. Evolutionary Computation, Vol. 3, No. 2, 82-102, Jul. 1999.

    LOCAL Z, Xi AS EXT
    LOCAL i%
    Z = 0##
    FOR i% = 1 TO Nd%
        Xi = R(p%,i%,j&)
        Z  = Z + Xi^2 - 10##*COS(TwoPi*Xi) + 10##
    NEXT i%
    Rastrigin = -Z
END FUNCTION 'Rastrigin
```

```
'----------------------
FUNCTION SGO(R(),Nd%,p%,j&)

'SGO Function (2-D)

'MAXIMUM = ~130.8323226... @ ~(-2.8362075...,-2.8362075...) WITH ZERO OFFSET.

'Reference: Hsiao, Y., Chuang, C., Jiang, J., and Chien, C., "A Novel Optimization Algorithm: Space
'Gravitational Optimization," Proc. of 2005 IEEE International Conference on Systems, Man,
'and Cybernetics, 3, 2323-2328. (2005)
    LOCAL x1, x2, Z, t1, t2, SGOx1offset, SGOx2offset AS EXT

    SGOx1offset = 0## : SGOx2offset = 0##

    x1 = R(p%,1,j&) - SGOx1offset : x2 = R(p%,2,j&) - SGOx2offset

    t1 = x1^4 - 16##*x1^2 + 0.5##*x1 : t2 = x2^4 - 16##*x2^2 + 0.5##*x2

    Z = t1 + t2

    SGO = -Z 'for maximization
END FUNCTION 'SGO()

'-----------------
FUNCTION GP(R(),Nd%,p%,j&)

'MAXIMUM = -3 @ (0,-1) WITH ZERO OFFSET.

'Reference: Cui, Z., Zeng, J., and Sun, G. (2006) 'A Fast Particle Swarm Optimization,' Int'l. J.
'Innovative Computing, Information and Control, vol. 2, no. 6, December, pp. 1365-1380.
    LOCAL Z, x1, x2, offset1, offset2, t1, t2 AS EXT

    offset1 = 0## : offset2 = 0##
'    offset1 = 20## : offset2 = -10##

    x1 = R(p%,1,j&)-offset1 : x2 = R(p%,2,j&)-offset2

    t1 = 1##+(x1+x2+1##)^2*(19##-14##*x1+3##*x1^2-14##*x2+6##*x1*x2+3##*x2^2)

    t2 = 30##+(2##*x1-3##*x2)^2*(18##-32##*x1+12##*x1^2+48##*x2-36##*x1*x2+27##*x2^2)

    Z  = t1*t2

    GP = -Z
END FUNCTION 'GP()

'------------------
FUNCTION CosineMix(R(),Nd%,p%,j&) '(n-D)

'MAXIMUM = 0.1*Nd @ [0]^Nd (n-D CASE).

'Reference: Pehlivanoglu, Y. V., "N New Particle Swarm Optimization Method Enhanced With a
'Periodic Mutation Strategy and Neural Networks," IEEE Trans. Evol. Computation,
'Vol. 17, No. 3, June 2013, pp. 436-452.
    LOCAL Z, Xi, Sum1, Sum2 AS EXT

    LOCAL i%

    Z = 0## : Sum1 = 0## : Sum2 = 0##

    FOR i% = 1 TO Nd%

        Xi   = R(p%,i%,j&)

        Sum1 = Sum1 + Xi^2

        Sum2 = Sum2 + COS(FivePi*Xi)

    NEXT i%

    Z = Sum1 - 0.1##*Sum2

    COSINEMIX = -Z
END FUNCTION 'COSINEMIX

'---------------------
FUNCTION Exponential(R(),Nd%,p%,j&) '(n-D)
'MAXIMUM = 1 @ [0]^Nd (n-D CASE).
'Reference: Pehlivanoglu, Y. V., "N New Particel Swarm Optimization Method Enhanced With a
'Periodic Mutation Strategy and Neural Networks," IEEE Trans. Evol. Computation,
'Vol. 17, No. 3, June 2013, pp. 436-452.
    LOCAL Z, Xi, Sum AS EXT

    LOCAL i%

    Z = 0## : Sum = 0##

    FOR i% = 1 TO Nd%

        Xi   = R(p%,i%,j&)

        Sum = Sum + Xi^2

    NEXT i%

    Z = -EXP(-0.5##*Sum)

    EXPONENTIAL = -Z
END FUNCTION 'EXPONENTIAL

'-----------------------
FUNCTION Ackley(R(),Nd%,p%,j&) '(n-D)

'MAXIMUM = ZERO (n-D CASE).

'References: (1) Yao, X., Liu, Y., and Lin, G., "Evolutionary Programming Made Faster,"
'IEEE Trans. Evolutionary Computation, Vol. 3, No. 2, 82-102, Jul. 1999.

'(2) Pehlivanoglu, Y. V., "N New Particel Swarm Optimization Method Enhanced With a
'Periodic Mutation Strategy and Neural Networks," IEEE Trans. Evol. Computation,
'Vol. 17, No. 3, June 2013, pp. 436-452.
    LOCAL Z, Xi, Sum1, Sum2 AS EXT
```

```
    LOCAL i%

    Z = 0## : Sum1 = 0## : Sum2 = 0##

    FOR i% = 1 TO Nd%

        Xi    = R(p%,i%,j&)

        Sum1 = Sum1 + Xi^2

        Sum2 = Sum2 + COS(TwoPi*Xi)

    NEXT i%

    Z = -20##*EXP(-0.2##*SQR(Sum1/Nd%)) - EXP(Sum2/Nd%) + 20## + e

    ACKLEY = -Z

END FUNCTION 'ACKLEY
'------------------

FUNCTION Griewank(R(),Nd%,p%,j&) 'Griewank (n-D)

'Max of zero at (0,...,0)

'References: (1) Yao, X., Liu, Y., and Lin, G., "Evolutionary Programming Made Faster,"
'IEEE Trans. Evolutionary Computation, Vol. 3, No. 2, 82-102, Jul. 1999.
'(2) Pehlivanoglu, Y. V., "N New Particel Swarm Optimization Method Enhanced With a
'Periodic Mutation Strategy and Neural Networks," IEEE Trans. Evol. Computation,
'Vol. 17, No. 3, June 2013, pp. 436-452.

    LOCAL Offset, Sum, Prod, Z, Xi AS EXT

    LOCAL i%

    Sum = 0## : Prod = 1##

    Offset = 0## '75.123##

    FOR i% = 1 TO Nd%

        Xi    = R(p%,i%,j&) - Offset

        Sum = Sum + Xi^2

        Prod = Prod*COS(Xi/SQR(i%))

    NEXT i%

    Z = Sum/4000## - Prod + 1##

    Griewank = -Z

END FUNCTION 'Griewank()
'-------------------------

FUNCTION Colville(R(),Nd%,p%,j&) 'Colville Function (4-D)

'MAXIMUM = 0 @ (1,1,1,1) WITH ZERO OFFSET.

'Reference: Doo-Hyun, and Se-Young, O., "A New Mutation Rule for Evolutionary Programming Motivated
'from Backpropagation Learning," IEEE Trans. Evolutionary Computation, Vol. 4, No. 2, pp. 188-190,
'July 2000.

    LOCAL Z, x1, x2, x3, x4, offset AS EXT

    offset = 0## '7.123##

    x1 = R(p%,1,j&)-offset : x2 = R(p%,2,j&)-offset : x3 = R(p%,3,j&)-offset : x4 = R(p%,4,j&)-offset

    Z =  100##*(x2-x1^2)^2 + (1##-x1)^2  + _
          90##*(x4-x3^2)^2 + (1##-x3)^2  + _
         10.1##*((x2-1##)^2 + (x4-1##)^2) + _
         19.8##*(x2-1##)*(x4-1##)

    Colville = -Z

END FUNCTION 'Colville()
'----------------------

'Note: NumPiFractions&&, PiFractions() are GLOBAL
FUNCTION RandomNumPiFrac(a,b,k&&) 'Returns random number X, a=< X < b.
    RandomNumPiFrac = a + (b-a)*PiFractions(k&&)
END FUNCTION 'RandomNumPiFrac()

'-----------------------------

FUNCTION RandomIntegerPiFrac%(n%,m%,k&&) 'Returns random integer I, n=< I =< m.
    RandomIntegerPiFrac = n% + (m%-n%)*PiFractions(k&&)
END FUNCTION 'RandomIntegerPiFrac%()

'---------------------------------

SUB MathematicalConstants
    EulerConst = 0.577215664901532860606512##
    Pi         = 3.141592653589793238462643##
    Pi2        = Pi/2##
    Pi4        = Pi/4##
    TwoPi      = 2##*Pi
    FourPi     = 4##*Pi
    FivePi     = 5##*Pi
    e          = 2.718281828459045235360287##
    Root2      = 1.414213562373095048##
END SUB
'------

SUB AlphabetAndDigits
    Alphabet$ = "ABCDEFGHIJKLMNOPQRSTUVWXYZabcdefghijklmnopqrstuvwxyz"
    Digits$   = "0123456789"
    RunID$    = REMOVE$(DATE$+TIME$,ANY" -:/")
END SUB
'------

SUB SpecialSymbols
    Quote$            = CHR$(34) 'Quotation mark "
    SpecialCharacters$ = "'(),#:;/_"
END SUB
'-----

SUB EMconstants
    Mu0  = 4E-7##*Pi        'hy/meter
```

```
    Eps0 = 8.854##*1E-12 'fd/meter
    c    = 2.998E8##     'velocity of light, 1##/SQR(Mu0*Eps0) 'meters/sec
    eta0 = SQR(Mu0/Eps0) 'impedance of free space, ohms
END SUB

'------

SUB ConversionFactors
    RadToDeg      = 180##/Pi
    DegToRad      = 1##/RadToDeg
    Feet2Meters   = 0.3048##
    Meters2Feet   = 1##/Feet2Meters
    Inches2Meters = 0.0254##
    Meters2Inches = 1##/Inches2Meters
    Miles2Meters  = 1609.344##
    Meters2Miles  = 1##/Miles2Meters
    NautMi2Meters = 1852##
    Meters2NautMi = 1##/NautMi2Meters
END SUB

'---------------------------- END PROGRAM GASR_PI_FRAC_01-13-2014.BAS ----------------------------
```